\documentstyle[twocolumn,aps,prl,epsf,rotate]{revtex}
\tightenlines

\def\Ps56{{\cal P}_Z(s_{56})}
\def\A60{{\cal A}^{{\rm tree}}_6}
\def\d{{\rm d}}

\def\res#1#2#3{$\!\!(#1\pm #2)\cdot 10^{#3}\!\!$}

\newcommand\as{{\alpha_s}}
\newcommand\ep{{\varepsilon}}
\newcommand\NLO{next-to-leading order }

\newcommand\qb{{\bar q}}
\newcommand\Qb{{\bar Q}}

\newcommand{\Nf}{{N_f}}
\newcommand\M{{\cal M}}

\newcommand\Tr{{\rm Tr}}

\newcommand{\beq}{\begin{equation}}
\newcommand{\eeq}{\end{equation}}
\newcommand{\beqn}{\begin{eqnarray}}
\newcommand{\eeqn}{\end{eqnarray}}
\newcommand\nn{\nonumber}

\begin{document}

\onecolumn
\preprint{hep-ph/9707309}
\title{Next-to-Leading Order Calculation of Four-Jet Shape Variables}
\author{Zolt\'an Nagy$^a$ and Zolt\'an Tr\'ocs\'anyi$^{b,a}$}
\address{$^a$Department of Theoretical Physics, KLTE,
H-4010 Debrecen P.O.Box 5, Hungary
\\
$^b$Institute of Nuclear Research of the Hungarian Academy of Sciences,
H-4001 Debrecen P.O.Box 51, Hungary}
\date{\today}
\maketitle

\begin{abstract}
We present the \NLO calculation of two four-jet event shape variables,
the D parameter and acoplanarity differential distributions.
We find large, more than 100\,\% radiative corrections. The theoretical
prediction  for the D parameter is compared to L3 data obtained at the
$Z^0$ peak and corrected to hadron level.
\end{abstract}



\twocolumn
In the second phase of the Large Electron Positron Collider
it is an important question how well the
characteristics of QCD four-jet events, i.e.\ events in which an
$s$-channel $Z^0$ or $\gamma^*$ decays into four quark and gluon jets,
are understood at large energies. This question is of interest because
$W^+W^-$ events lead to four-jet final states for which the main
backgrounds are QCD events and because QCD four-jet events are also the
principal source of background for Higgs and other new particle
searches. The perturbative description of the QCD four-jet
events is also interesting in its own right as a tool for testing
perturbation theory in a regime with small hadronization uncertainty
and for measuring the QCD color charges \cite{LEP}, or as a
means of testing whether experimental data favor or exclude the
existence of light gluinos \cite{gluino}.

Recent theoretical developments make possible the \NLO calculation of
four-jet quantities. There are now several general methods available
for the cancellation of infrared divergences that can be used for setting
up a Monte Carlo evaluation of \NLO partonic cross sections
\cite{slicing,residue,dipole}. The main ingredients of the calculation
are the four-parton \NLO and five-parton Born level squared matrix
elements. The tree level amplitudes for the processes $e^+e^-\to \qb q
ggg$ and $e^+e^-\to \qb q \Qb Q g$ from which the latter can be
constructed have been known for a long time \cite{a5parton}.
Recently Campbell, Glover and Miller calculated the other vital piece
of information, the virtual corrections for the processes
$e^+e^-\to \gamma^*\to \qb q \Qb Q$ and $\qb q g g$ \cite{CGM}. Also,
the new techniques developed by Bern, Dixon and Kosower in the
calculation of one-loop multiparton helicity amplitudes
\cite{BDKannals} made possible the derivation of analytic expressions
for the helicity amplitudes of the $e^+e^-\to Z^0,\,\gamma^*\to \qb q
\Qb Q$ process \cite{BDKW4q} and results for the other subprocess are
expected to appear soon. Using these results Dixon and Signer
calculated the \NLO corrections for four-jet fractions
with various clustering algorithms \cite{menloparc,DSjets}.

In this Letter we enlarge the list of four-jet observables that are
calculated to \NLO accuracy. We present results of the
calculation of QCD radiative corrections to two four-jet shape variable
differential distributions --- the D parameter and acoplanarity.

We use the matrix elements of Ref.~\cite{CGM} for the loop corrections.
In the calculation of these matrix elements all quark and lepton massess
are set to zero, thus our results are valid in the massless limit.
We note that the results in Ref.~\cite{CGM} do not include the
``light-by-glue'' virtual contributions which were shown
to be negligible \cite{DSjets}.


The higher order correction to the leading order partonic cross section
$\sigma^{\rm LO}$ is a sum of two integrals, one of the real correction
$\d\sigma^{\rm R}$ that is an exclusive cross section of five partons
in the final state and the other of the virtual correction
$\d\sigma^{\rm V}$ that is the one-loop correction to the process with
four partons in the final state:
\beq
\label{sNLO}
\sigma^{\rm NLO} \equiv \int \d\sigma^{\rm NLO}
= \int_5 \d\sigma^{\rm R}
+ \int_4 \d\sigma^{\rm V}\:,
\eeq                                                                 
where
\beq
\int_5 \d\sigma^{\rm R}
= \int \d\Gamma^{(5)} <|\M_5^{\rm tree}|^2> J_5
\eeq                                                                 
and 
\beq
\int_4 \d\sigma^{\rm V}
= \int \d\Gamma^{(4)} <|\M_4^{\rm 1-loop}|^2> J_4 \:.
\eeq                                                                 

The two integrals on the right-hand side of Eq.~(\ref{sNLO}) are separately
divergent in $d=4$ dimensions, but their sum is finite provided the jet
function $J_n$ defines an infrared safe quantity. Therefore, the separate
pieces have to be regularized. We use dimensional regularization in
$d=4-2\ep$ dimensions, in which case the divergences are replaced by 
double and single poles in $\ep$. We assume that ultraviolet
renormalization of all Green functions to one-loop order has been
carried out, so the poles are of infrared origin.

There are several ways of exposing the cancellation of infrared singularities
directly at the integrand level \cite{slicing,residue,dipole}. The
method used in the calculation presented in this Letter is a slightly
modified version of the dipole formalism of Catani and Seymour
\cite{dipole} that is based on the subtraction method. The general
idea of the subtraction method for writing a general-purpose Monte
Carlo program is to use the identity
\beq
\label{sNLO2}
\sigma^{\rm NLO} =
\int_5 \left[\d\sigma^{\rm R} - \d\sigma^{\rm A}\right]
+ \int_4 \left[ \d\sigma^{\rm V} + \int_1 \d\sigma^{\rm A} \right] \:,
\eeq
where $\d\sigma^{\rm A}$ in the dipole formalism is a proper
approximation of $\d\sigma^{\rm R}$ in the kinematically degenerate
(soft and collinear) region so that it has the same pointwise singular
behaviour (in $d$ dimensions) as $\d\sigma^{\rm R}$ itself, As a
result, $\d\sigma^{\rm A}$ acts as a local counterterm for
$\d\sigma^{\rm R}$, that is, $[\d\sigma^{\rm R} - \d\sigma^{\rm A}]$
is integrable in four dimensions by definition. The approximate cross
section is constructed in such a way that it can be integrated
analytically over the exactly factorized one-parton subspace leading
to $\ep$ poles, that can be combined with those in $\d\sigma^{\rm V}$.
The $\ep$ poles are guaranteed to cancel for infrared safe observables
(Kinoshita-Lee-Nauenberg theorem).  These quantities
have to be experimentally (theoretically) defined in such a way that
their actual value is independent of the number of soft and collinear
hadrons (partons) produced in the final state. In particular, this   
value has to be the same in a given four-parton configuration and in all
five-parton configurations that are kinematically degenerate with it
(i.e.\ that are obtained from the four-parton configuration by adding a
soft parton or replacing a parton with a pair of collinear partons
carrying the same total momentum). This property can be simply
restated in a formal way. If the function $J_{n}$ gives the value of
a certain jet observable in terms of the momenta of the $n$ final-state
partons, we should have
\beq\label{Jn}
J_5 \to J_4 \:,
\eeq
in any case where the five-parton and the four-parton configurations
are kinematically degenerate. It is easy to prove that the observables
considered in this Letter fulfill this property. When the requirement
of infrared safety, relation (\ref{Jn}) is fulfilled the second integral in
Eq.~(\ref{sNLO2}) is also finite in $d=4$ dimensions and $\sigma^{\rm NLO}$
can be easily implemented in a `partonic Monte Carlo' program that
generates appropriately weighted partonic events with five final-state
partons and events with four partons.

For the precise definition of the approximate cross section in the dipole
formalism, we refer to the original work of Catani and Seymour
\cite{dipole}. The distinct feature of this formalism as compared to
other subtraction methods \cite{residue} is the exact factorization of
the five-particle phase space into a four-particle and a one-particle
phase space, and that the approximate cross sections provides a single
and smooth approximation of the real cross section in all of its
singular limits. These features lead to a well-converging partonic
Monte Carlo program.


In this Letter we consider two classic four-jet event shape variables. 
The D parameter \cite{Dpar} is derived from the eigenvalues of the
infrared safe momentum tensor 
\beq
\theta^{ij} =
\sum_a \frac{p_a^i p_a^j}{|\vec{p}_a|}\bigg/\sum_a |\vec{p}_a|,
\eeq
where the sum on $a$ runs over all final state hadrons and $p_a^i$ is the
$i$th component of the three-momentum $\vec{p}_a$ of hadron $a$ in the
c.m. system. The tensor $\theta$ is normalized to have unit trace. In
terms of the eigenvalues $\lambda_i$ of the $3 \times 3$ matrix
$\theta$, the global shape parameter D is defined as
\beq\label{Ddef}
{\rm D} = 27\,\lambda_1 \lambda_2 \lambda_3\:.
\eeq

The second observable is acoplanarity \cite{acop} defined as
\beq
\label{Adef}
{\rm A} = 4\min
\left(\frac{\sum_a |\vec{p}_a^{\rm out}|}{\sum_a|\vec{p}_a|}\right)^2,
\eeq
where the sum runs over all particles in an event, and $\vec{p}_a^{\rm
out}$ is measured perpendicular to a plane chosen to minimize A.

In the case of the shape variable differential distributions for
observable $O$ the jet function $J_n$ is actually a functional,
\beq
J_n = \delta(O-O^{(n)})\:,
\eeq
where D$^{(n)}$ is given by Eq.~(\ref{Ddef}) and A$^{(n)}$ is given by
Eq.~(\ref{Adef}).


Once the integrations in Eq.~(\ref{sNLO2}) are carried out, the \NLO
differential cross section for the four-jet observable $O_4$ takes the
general form
\beqn
&&\frac{1}{\sigma_0}O_4\frac{\d \sigma}{\d O_4}(O_4)
= \left(\frac{\as(\mu)}{2\pi}\right)^2 B_{O_4}(O_4)
\\ \nn &&\qquad
+ \left(\frac{\as(\mu)}{2\pi}\right)^3
\left[B_{O_4}(O_4)\beta_0\ln\frac{\mu^2}{s} + C_{O_4}(O_4)\right]\:.
\eeqn
In this equation $\sigma_0$ denotes the Born cross section for the
process $e^+e^-\to \qb q$,
$\beta_0 = \left(\frac{11}{3}C_A - \frac{4}{3} T_R \Nf\right)$
with the normalization $T_R=\frac{1}{2}$ in
$\Tr(T^aT^{\dag b})=T_R\delta^{ab}$, $s$ is the total c.m.\ energy
squared, $\mu$ is the renormalization scale, while $B_{O_4}$ and
$C_{O_4}$ are scale independent functions, $B_{O_4}$ is the Born
approximation and $C_{O_4}$ is the radiative correction.

The first complete results obtained for four-jet observables at
\NLO accuracy \cite{DSjets} are four-jet rates for three clustering
algorithms: the Durham~\cite{durham}, the Geneva~\cite{geneva} and the
E0~\cite{E0} schemes calculated for three colors, five massless
flavors and with $\as(M_Z)=0.118$. We also calculated these observables
and compared the results of the two calculations in Table~I. 
There is a very good agreement between the two calculations apart from
the 3\,\% discrepancy in the Geneva scheme result at $y_{\rm cut}=0.05$.
\vbox{
\begin{table}             
\caption{Comparison of the four-jet fractions calculated by the two
partonic Monte Carlo programs MENLO PARC and DEBRECEN (this work).}
\begin{tabular}{lccc}
Algorithm&$y_{\rm cut}$&  MENLO PARC         &   DEBRECEN          \\   
\tableline                        
         &    0.005    &\res{1.04}{0.02}{-1} &\res{1.05}{0.004}{-1}\\
 Durham  &    0.01     &\res{4.70}{0.06}{-2} &\res{4.66}{0.02}{-2} \\
         &    0.03     &\res{6.82}{0.08}{-3} &\res{6.87}{0.04}{-3} \\
\tableline                                                          
         &    0.02     &\res{2.56}{0.06}{-1} &\res{2.63}{0.06}{-1} \\
 Geneva  &    0.03     &\res{1.71}{0.03}{-1} &\res{1.75}{0.03}{-1} \\
         &    0.05     &\res{8.58}{0.15}{-2} &\res{8.27}{0.08}{-2} \\
\tableline              
         &   0.005     &\res{3.79}{0.08}{-1} &\res{3.88}{0.07}{-1} \\
   E0    &    0.01     &\res{1.88}{0.03}{-1} &\res{1.92}{0.01}{-1} \\
         &    0.03     &\res{3.46}{0.05}{-2} &\res{3.37}{0.01}{-2} \\
\end{tabular}                                               
\end{table}
\indent
We list the numerical values for $B_{\rm D}$, $C_{\rm D}$ in Table~II
and those for $B_{\rm A}$ and $C_{\rm A}$ in Table~III. Our program
generates four and five parton events with an appropiate weight. In
order to obtain the $B_{O_4}$ and $C_{O_4}$ functions we calculated the
$O_4$ observable of each event, multiplied each weight by $O_4$ and
added to the appropiate bin $O_4$.

We define the average value of these shape variables as
\beq
<O_4> = \frac{1}{\sigma} \int_0^1 \d O_4\,O_4\frac{\d \sigma}{\d O_4}\:.
\eeq
We studied the dependence of the average value of the D parameter on
the renormalization scale in Fig.~1. The strong dependence found at
leading order is decreased at next-to-leading order. However, there
still remains substantial scale dependence showing that the
uncalculated higher order corrections are presumably large. The feature
is similar in the case of acoplanarity, but the residual scale dependence
is even larger.
}
\begin{table}
\caption{The Born level and \NLO scale independent functions $B_{\rm D}$
and $C_{\rm D}$.}
\begin{tabular}{lll}
   D   &      $B_{\rm D}$     &       $C_{\rm D}$    \\
\tableline
  0.00 & \res{6.60}{0.02}{ 2} & \res{1.08}{0.06}{ 4} \\
  0.04 & \res{2.32}{0.01}{ 2} & \res{1.24}{0.02}{ 4} \\
  0.08 & \res{1.45}{0.01}{ 2} & \res{8.59}{0.12}{ 3} \\
  0.12 & \res{1.03}{0.01}{ 2} & \res{6.24}{0.12}{ 3} \\
  0.16 & \res{7.74}{0.05}{ 1} & \res{4.99}{0.11}{ 3} \\
  0.20 & \res{5.97}{0.04}{ 1} & \res{3.85}{0.06}{ 3} \\
  0.24 & \res{4.69}{0.03}{ 1} & \res{2.98}{0.05}{ 3} \\
  0.28 & \res{3.77}{0.03}{ 1} & \res{2.52}{0.05}{ 3} \\
  0.32 & \res{3.01}{0.02}{ 1} & \res{1.94}{0.05}{ 3} \\
  0.36 & \res{2.41}{0.02}{ 1} & \res{1.59}{0.04}{ 3} \\
  0.40 & \res{1.98}{0.02}{ 1} & \res{1.37}{0.03}{ 3} \\
  0.44 & \res{1.61}{0.02}{ 1} & \res{1.06}{0.03}{ 3} \\
  0.48 & \res{1.30}{0.01}{ 1} & \res{8.72}{0.19}{ 2} \\
  0.52 & \res{1.07}{0.01}{ 1} & \res{7.11}{0.16}{ 2} \\
  0.56 & \res{8.48}{0.10}{ 0} & \res{5.68}{0.14}{ 2} \\
  0.60 & \res{6.70}{0.09}{ 0} & \res{4.46}{0.21}{ 2} \\
  0.64 & \res{5.33}{0.08}{ 0} & \res{3.52}{0.11}{ 2} \\
  0.68 & \res{4.10}{0.07}{ 0} & \res{2.74}{0.09}{ 2} \\
  0.72 & \res{3.11}{0.06}{ 0} & \res{2.08}{0.08}{ 2} \\
  0.76 & \res{2.24}{0.05}{ 0} & \res{1.54}{0.06}{ 2} \\
  0.80 & \res{1.52}{0.04}{ 0} & \res{1.03}{0.04}{ 2} \\
  0.84 & \res{9.95}{0.30}{-1} & \res{6.66}{0.31}{ 1} \\
  0.88 & \res{5.74}{0.22}{-1} & \res{3.89}{0.20}{ 1} \\
  0.92 & \res{2.68}{0.15}{-1} & \res{1.71}{0.19}{ 1} \\
  0.96 & \res{5.16}{0.61}{-2} & \res{2.60}{1.30}{ 0} \\
\end{tabular} 
\end{table}
\begin{table}
\caption{The Born level and \NLO scale independent functions $B_{\rm A}$
and $C_{\rm A}$.}
\begin{tabular}{lll}
   A   &      $B_{\rm A}$     &       $C_{\rm A}$    \\
\tableline
  0.00 & \res{3.34}{0.01}{ 2} & \res{1.56}{0.01}{ 4} \\
  0.04 & \res{7.39}{0.03}{ 1} & \res{5.17}{0.08}{ 3} \\
  0.08 & \res{3.63}{0.02}{ 1} & \res{2.69}{0.06}{ 3} \\
  0.12 & \res{2.05}{0.01}{ 1} & \res{1.56}{0.03}{ 3} \\
  0.16 & \res{1.23}{0.01}{ 1} & \res{9.59}{0.22}{ 2} \\
  0.20 & \res{7.63}{0.07}{ 0} & \res{6.12}{0.15}{ 2} \\
  0.24 & \res{4.81}{0.05}{ 0} & \res{3.97}{0.12}{ 2} \\
  0.28 & \res{3.02}{0.04}{ 0} & \res{2.57}{0.09}{ 2} \\
  0.32 & \res{1.78}{0.03}{ 0} & \res{1.59}{0.08}{ 2} \\
  0.36 & \res{1.08}{0.02}{ 0} & \res{1.10}{0.06}{ 2} \\
  0.40 & \res{5.99}{0.17}{-1} & \res{5.99}{0.33}{ 1} \\
  0.44 & \res{3.19}{0.12}{-1} & \res{3.67}{0.25}{ 1} \\
  0.48 & \res{1.51}{0.09}{-1} & \res{1.90}{0.99}{ 1} \\
  0.52 & \res{5.91}{0.52}{-2} & \res{8.45}{0.99}{ 0} \\
  0.56 & \res{1.55}{0.26}{-2} & \res{2.84}{0.42}{ 0} \\
  0.60 & \res{1.33}{0.84}{-3} & \res{7.64}{1.77}{-1} \\
  0.64 & \res{0.00}{0.00}{ 0} & \res{5.55}{2.66}{-2} \\
\end{tabular} 
\end{table}

\begin{figure}
\epsfxsize=5cm \begin{rotate}[r]{\epsfbox{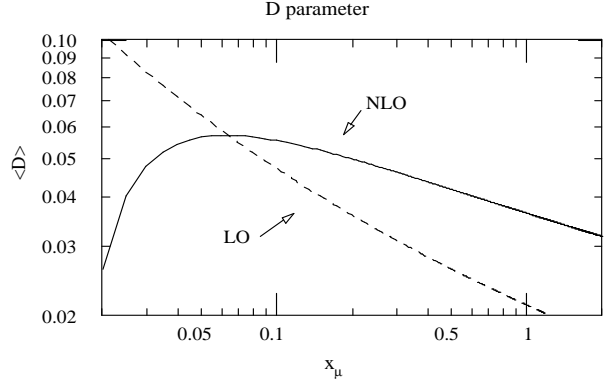}}\end{rotate}
\caption{Renormalization scale dependence of the average value of the D
parameter. $x_\mu = \mu/\sqrt{s}$}
\end{figure}

The same conclusion is drawn if we look at the dependence of the K factors 
on the observables as depicted in Fig.~2. In case of the D parameter
the K factor is slightly above two for the whole range, while for
acoplanarity it is even larger and increases for larger values of A.
This suggests that A cannot be reliable calculated in perturbation
theory.

Finally, in Fig.~3 we compare the \NLO QCD prediction for the D
parameter to L3 data obtained at the $Z^0$ peak \cite{L3Dpar} and
corrected to hadron level. The inclusion of the higher order correction
decreases the discrepancy between the \NLO QCD prediction and the data.
However, there still remains significant discrepancy. This difference
may come in part from hadronization effects, and also from the
uncalculated even higher order contributions.

\begin{figure}
\epsfxsize=7cm \begin{rotate}[r]{\epsfbox{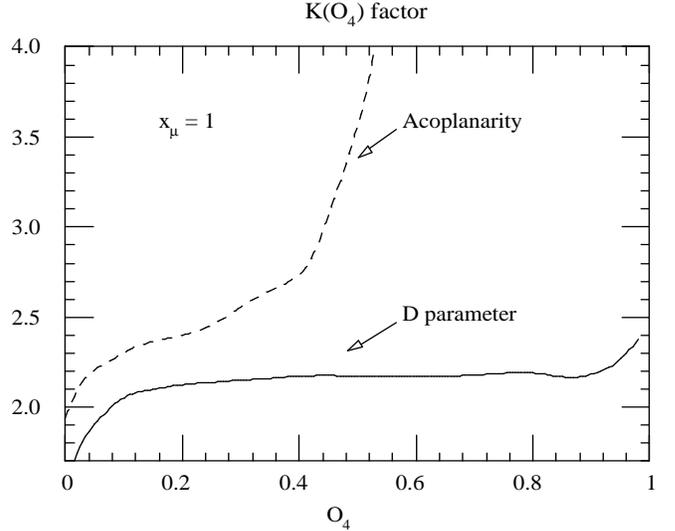}}\end{rotate}
\caption{K factor of the D parameter and acoplanarity.}
\end{figure}

\begin{figure}
\epsfxsize=7.5cm \begin{rotate}[r]{\epsfbox{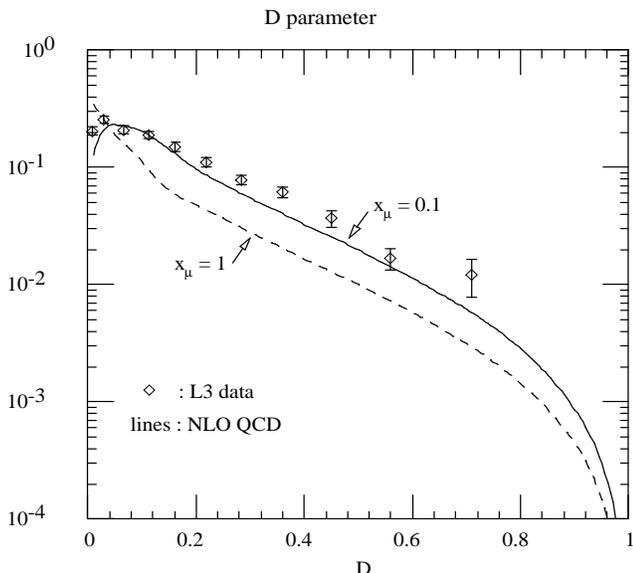}}\end{rotate}
\caption{Comparison of the \NLO QCD prediction for the D parameter
differential distribution,
$\frac{{\rm D}}{\sigma}\frac{\d\sigma}{\d {\rm D}}$
to L3 data obtained at the $Z^0$ peak and corrected to hadron level.
The upper edge of the theoretical band is obtained with renormalization
scale $x_\mu=0.1$, while the lower edge at $x_\mu = 1$. }
\end{figure}


In this Letter we presented for the first time a \NLO calculation of
the differential cross section of two classic four jet shape variables,
the D parameter and acoplanarity. We gave explicit results for the
radiative corrections to the leading order cross sections. The
corrections are large indicating that the uncalculated even higher order
terms are important. This feature is especially dramatic in the case of
acomplanarity suggesting that this observable cannot be reliable
calculated in perturbation theary. We also compared the
four-jet rates obtained by our program to the results of Dixon and
Signer \cite{DSjets} and found agreement.

These results were produced by a partonic Monte Carlo program
that can be used for the calculation of QCD radiative corrections to
the differential cross section of any kind of four-jet observable
in electron-positron annihilation.

This research was supported in part by the EEC Programme "Human Capital
and Mobility", Network "Physics at High Energy Colliders", contract
PECO ERBCIPDCT 94 0613 as well as by the Hungarian Scientific Research
Fund grant OTKA T-016613 and the Research Group in Physics of the
HAS, Debrecen.

{\bf Note added:} After the completion of this work, the helicity
amplitudes of the $e^+e^-\to  Z^0,\,\gamma^*\to \qb q g g$ process
have been published \cite{BDK2q2g}, and the agreement with the
results of Ref.~\cite{CGM} in the $\gamma^*$ channel has been established.
Also, the authors of Ref.~\cite{DSjets} pointed out a slight error in our
binning procedure in the case of the Geneva algortihm. Correcting this
error we find $R_4(y_{\rm cut} = 0.05) = (8.37\pm0.12)\cdot 10^{-2}$ that
agrees with the result of Ref.~\cite{DSjets}.

\def\np#1#2#3  {Nucl.\ Phys.\ {\bf #1}, #2 (19#3)}
\def\pl#1#2#3  {Phys.\ Lett.\ {\bf #1}, #2 (19#3)}
\def\prep#1#2#3  {Phys.\ Rep.\ {\bf #1}, #2 (19#3)}
\def\prd#1#2#3 {Phys.\ Rev.\ D {\bf #1}, #2 (19#3)}
\def\prl#1#2#3 {Phys.\ Rev.\ Lett.\ {\bf #1}, #2 (19#3)}
\def\zpc#1#2#3  {Zeit.\ Phys.\ C {\bf #1}, #2 (19#3)}
\def\cmc#1#2#3  {Comp.\ Phys.\ Comm.\ {\bf #1}, #2 (19#3)}
\def\anr#1#2#3  {Ann.\ Rev.\ Nucl.\ Part.\ Sci.\ #1 (19#3) #2}

\end{document}